\documentclass[a4paper,11pt]{article}
\usepackage{jinstpub} 
\usepackage{lineno}
\usepackage{caption}
\usepackage{subcaption}
\usepackage{comment}
\graphicspath{{figures/}}
\usepackage{xcolor}
\usepackage{soul}

\title{\boldmath First results on monolithic CMOS detector with internal gain}

\author[a,b,1]{U. Follo\note[1]{\label{author}Corresponding authors.},}
\author[a,c,1]{G. Gioachin,}
\author[a,b]{C. Ferrero,}
\author[a]{M. Mandurrino,}
\author[d]{M. Bregant,}
\author[a,c]{S. Bufalino}
\author[e]{F. Carnesecchi,}
\author[f]{D. Cavazza,}
\author[f,g]{M. Colocci,}
\author[h,i,2]{T. Corradino\note[2]{Now at Fondazione Bruno Kessler},}
\author[a]{M. Da Rocha Rolo,}
\author[l]{G. Di Nicolantonio,}
\author[a,b]{S. Durando,}
\author[l]{G. Margutti,}
\author[a]{M. Mignone,}
\author[f]{R. Nania,}
\author[h,i]{L. Pancheri,}
\author[a]{A. Rivetti,}
\author[f,g]{B. Sabiu,}
\author[d]{G. G. A. de Souza,}
\author[f,g]{S. Strazzi,}
\author[a]{R. Wheadon}

\emailAdd{umberto.follo@polito.it}
\emailAdd{giulia.gioachin@polito.it}

\affiliation[a]{Istituto Nazionale di Fisica Nucleare (INFN), Sezione di Torino, Via P. Giuria 1, 10125, Torino, Italy}
\affiliation[b]{Dipartimento di Elettronica e Telecomunicazioni (DET), Politecnico di Torino, Corso Duca degli Abruzzi 24, 10129, Torino, Italy}
\affiliation[c]{Dipartimento di Scienza Applicata e Tecnologia (DISAT), Politecnico di Torino, Corso Duca degli Abruzzi 24, 10129, Torino, Italy}
\affiliation[d]{Instituto de F\'{i}sica da Universidade de S\~{a}o Paulo, Rua do Mat\~{a}o 1371, 05508-090 Cidade Universit\'{a}ria, S\~{a}o Paulo, Brazil}
\affiliation[e]{European Organisation for Nuclear Research (CERN), Esplanade des Particules 1, Geneva, 1211 Geneva 23, Switzerland}
\affiliation[f]{Istituto Nazionale di Fisica Nucleare (INFN), Sezione di Bologna, Viale C. Berti Pichat 6/2, 40127, Bologna, Italy}
\affiliation[g]{Dipartimento Fisica e Astronomia, Università di Bologna, Viale C. Berti Pichat 6/2, 40127, Bologna, Italy}
\affiliation[h]{Universit\`{a} di Trento, Via Sommarive 9, 38123, Trento, Italy}
\affiliation[i]{TIFPA-INFN, Via Sommarive 14, 38123, Trento, Italy}
\affiliation[l]{LFoundry S.r.l, Via Pacinotti 7, 67051, Avezzano (AQ), Italy}

\abstract{In this paper we report on a set of characterisations carried out on the first monolithic LGAD prototype integrated in a customised 110~nm CMOS process having a depleted active volume thickness of 48~$\mu$m. 
This prototype is formed by a pixel array where each pixel has a total size of 100~$\mu$m~$\times$~250~$\mu$m and includes a high-speed front-end amplifier.
After describing the sensor and the electronics architecture, both laboratory and in-beam measurements are reported and described.
Optical characterisations performed with an IR pulsed laser setup have shown a sensor internal gain of about 2.5.
With the same experimental setup, the electronic jitter was found to be between 50~ps and 150~ps, depending on the signal amplitude.
Moreover, the analysis of a test beam performed at the Proton Synchrotron (PS) T10 facility of CERN with 10~GeV/c protons and pions indicated that the overall detector time resolution is in the range of 234~ps to 244~ps.
Further TCAD investigations, based on the doping profile extracted from $C(V)$ measurements, confirmed the multiplication gain measured on the test devices.
Finally, TCAD simulations were used to tune the future doping concentration of the gain layer implant, targeting sensors with a higher avalanche gain. This adjustment is expected to enhance the timing performance of the sensors of the future productions, in order to cope with the high event rate expected in most of the near future high-energy and high-luminosity physics experiments, where the time resolution will be essential to disentangle overlapping events and it will also be crucial for Particle IDentification (PID).}

\keywords{Solid state detectors, Timing detectors, Front-end electronics for detector readout, Detector modelling and simulations II (electric fields, charge transport, multiplication and induction, pulse formation, electron emission, etc)}

\begin{document}
\maketitle
\flushbottom

\section{Introduction}
\label{sec:intro}
In past few years, CMOS Monolithic Active Pixel Sensors (MAPS) have widely demonstrated to meet the requirements of tracking detectors as a viable alternative to hybrid pixels in high-energy physics experiments \cite{REIDT2022166632, CONTIN201860, 2022Iacobucci_JINST}.
Since the spatial density of particle collisions expected in near-future high-energy physics experiments will dramatically increase, silicon detectors are requested to provide also precise time information to perform an accurate reconstruction of tracks~\cite{CMS,ATLAS}, particle identification~\cite{ALICE} or as beam-induced background mitigation technique (see, for instance, Ref.~\cite{MUCOL}).
Several experiments will face this problem, such as the next-generation heavy-ion experiment named ALICE~3~\cite{ALICE}, which will be installed at the Large Hadron Collider (LHC) at CERN during the long shutdown 4 (2033-2034).

Recently, significant research efforts have been focused on small pixel size, low noise, and low power consumption to develop MAPS for tracking applications, but their potential in terms of time resolution has not yet been fully exploited. Therefore, to achieve good timing performance a fast charge collection is necessary, and this requires often the use of fully depleted sensors. To this end, Fully Depleted MAPS (FD-MAPS) have been developed in the ARCADIA project~\cite{Pancheri_2020} where charge induction is due to the drift of carriers. 
However, this is still not sufficient to reach the optimal timing performance, as it is also needed to carefully optimise the device geometry and the related electric field shape. Sensors with small collection electrodes offer a small capacitance and a higher Signal-to-Noise Ratio (SNR), but the resulting drift paths have different lengths, limiting the time resolution. On the other hand, large collection electrodes show a better field uniformity with the drawback of a larger capacitance, and thus a decreased SNR. 
A promising solution to suppress the effect of noise increasing the SNR has been found by implanting a gain layer below the collection electrode, thus providing charge multiplication by impact ionization. In this way, the LGAD (Low Gain Avalanche Diode) technology~\cite{2014Pellegrini_NIMA} has been integrated in the present ARCADIA production of FD-MAPS devices produced with a commercial 110~nm CMOS Image Sensor (CIS) process provided by LFoundry.\\
Contrarily to traditional LGADs, which are produced on sensor-grade wafers and require dedicated readout electronics, monolithic sensors can enable to decrease the production costs and to simplify the complex assembly procedures characterising the hybrid devices. 

The innovative concept of this Monolithic CMOS Avalanche Detector PIXelated (MadPix) may find application not only in the detection of charged particles in high energy physics experiments but also in medical particle tomography \cite{IMPACT} and tracking in space experiments \cite{DeSantis:2021WO}, which would benefit from the low material budget, a fine pixel pitch, and a low power consumption. 

\hfill \break
In this article, we present the first prototype of a monolithic sensor with internal gain based on a commercial 110~nm CMOS process. In particular, Section 2 discusses the design of MadPix, covering the geometry and operation of the device, and the description of the integrated electronic and readout system. In Section 3, the laboratory results of the sensor characterisation are compared with the expectations of the performed simulations. In Section 4 we show the first test beam measurements and the relative results obtained at the Proton Synchrotron (PS) at CERN. Section 5 investigates the properties of the sensor through a numerical study based on a 2D TCAD framework explaining the MadPix behavior.
Finally, the conclusions and the perspectives for this technology are highlighted in Section 6. 

\section{Design}
\label{sec:design}
In this section, the description of the sensor layout and the front-end electronics is reported. Furthermore, an overview of the first prototype of monolithic CMOS sensor with internal gain in 110~nm, MadPix, is presented. Finally, the test-board for the chip readout is illustrated.
\subsection{Sensor}
\label{subsec:sensor}
The sensor concept is based on the fully-depleted MAPS developed within the ARCADIA project, where the add-on of a gain layer implant has been implemented~\cite{2024Corradino_JINST}.
In particular, MadPix design, together with a set of sensor test structures, were fabricated in the ARCADIA third engineering run.

\begin{figure}[!ht]
\centering
     \begin{subfigure}[!ht]{0.49\linewidth}
         \centering
         \includegraphics[width=\textwidth]{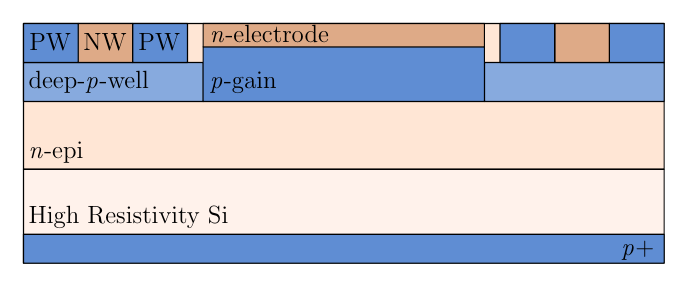}
         \caption{Layout A1}
         \label{fig:A1}
     \end{subfigure}
     \begin{subfigure}[!ht]{0.49\linewidth}
         \centering
         \includegraphics[width=\textwidth]{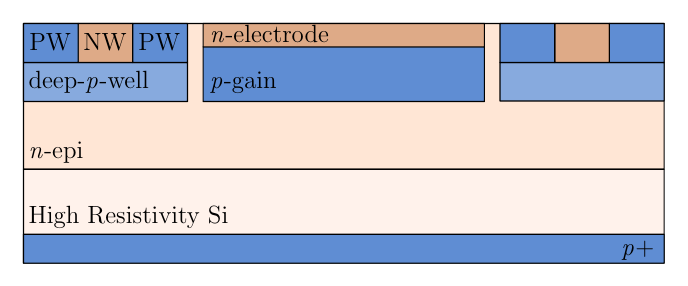}
         \caption{Layout A2}
         \label{fig:A2}
     \end{subfigure}
\caption{Cross-sectional view of the (a) A1 and (b) A2 termination layouts differing by the distance between the \textit{p}-gain and the deep-\textit{p}-well implants.}
\label{fig:layout}
\end{figure}

The devices are processed through the ion implantation technique on the chip frontside (see Figure~\ref{fig:layout}).
As in standard pixelated MAPS, a shallow \textit{n}$^+$ implant works as collection electrode, while the gain layer is obtained through a \textit{p}$^+$-type layer implanted just beneath the electrode.
Surrounding this \textit{n}$^+$ region, and the underlying gain layer, we find the deep-\textit{p}-wells hosting the front-end electronics (\textit{n}-wells, NW, for the PMOS and \textit{p}-wells, PW, for the NMOS).
These \textit{p}-type implants are crucial to achieve a full isolation between neighboring PMOS structures, as well as to prevent competitive charge collection that would take place among the NW and the \textit{n}$^+$ electrode.
Concerning the substrate, the upper part consists of a thin \textit{n}-type epitaxial layer followed by a second, slightly less doped, epitaxial layer of the same type.
These two regions represent the active volume of the sensor and have a global thickness of about 48~$\mu$m.
They are grown on a highly doped \textit{p}$^+$ silicon wafer, that completes the \textit{pn} junction of the device and allows to apply a reverse bias on the back contact.
The biases of the front and back contact are referred to the common ground (0~V) applied to the deep-\textit{p}-wells.
Typically, $-$40~V to $-$50~V are applied on the backside to have saturation of carriers velocity and operate in full depletion conditions.
While, regarding the frontside, applying a certain voltage on the \textit{n}$^+$ electrode (in the range between 30~V and 45~V) the electric field in the gain layer becomes high enough to trigger the onset of the charge multiplication through the impact ionization mechanism.
The higher the voltage applied to the collection electrode, the higher the gain. For what concerns the guard-rings, we usually bias them at 8-9~V since this voltage is sufficiently high to contribute, along with the backside bias, to the \mbox{full-depletion} of substrate, but at the same time it is safely low to avoid breakdown phenomena at the edges of the active region.

The ARCADIA project developed also fully depleted sensors with an active thickness of 100~$\mu$m and 200~$\mu$m starting from high resistivity \textit{n}-type substrates. However, only the 48~$\mu$m thick structures were analysed in this study because simulations and data have shown that thinner sensors lead to better time resolutions~\cite{2021Ferrero_CRC}.

Two different pixel layouts, differing by the termination at the pixel periphery, have been designed (see Figure~\ref{fig:layout}).
In the first one, called A1 (Figure~\ref{fig:A1}), the deep-\textit{p}-wells are in contact with the gain layer, determining an increased multiplication volume.
With this particular design, also primary charges generated via impact ionization under the \textit{p}-well are multiplied, after being drifted towards the gain layer and then collected by the frontside electrode.
However, these charges reach the \textit{n}$^+$ implant with a certain delay with respect to the ones produced within the active area, due to the field curvature experienced at pixel borders.
In the second layout, called A2 (Figure~\ref{fig:A2}), a small gap is provided between the two implants.
This means that some charges produced at border may follow drift lines that go through this gap and reach directly the \textit{n}$^+$ electrode without crossing the gain layer.
As a result, layout A2 allows the discrimination of signals generated in the region with gain, that are induced by charges whose drift lines are straightly directed to the collection electrode, from the peripheral ones, which are not multiplied.

\subsection{Electronics}
\label{subsec:preamplifier}
\begin{figure}[!ht]
\centerline{\includegraphics[width=\linewidth]{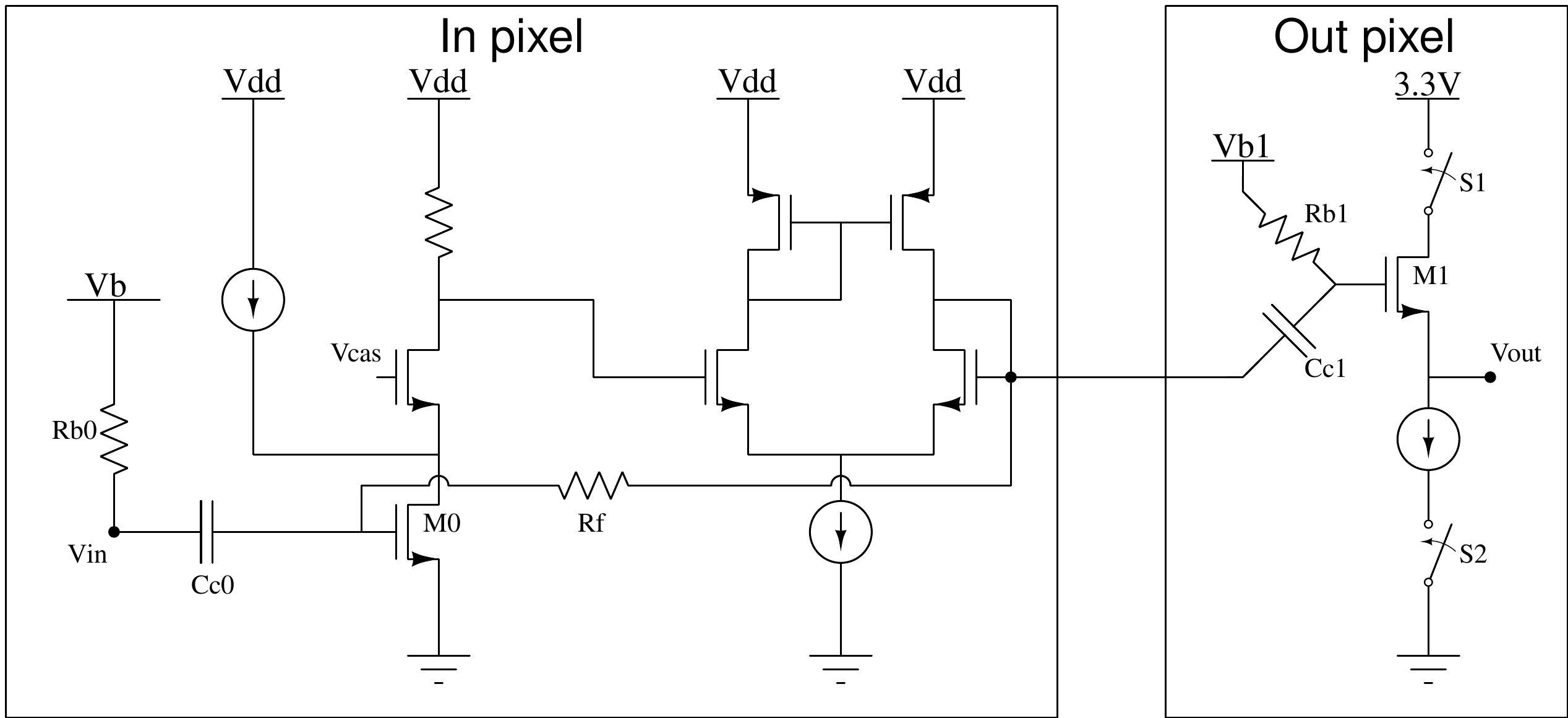}}
\caption{Schematic of the in pixel front-end electronics followed by the output buffer placed outside the matrices.}
\label{fig:schematic}
\end{figure}

The electronics of the chip is implemented in a 110~nm node and the amplifier design is based on a cascoded common source architecture, with bias split and active load (see Figure~\ref{fig:schematic}).
The input transistor (M0) is sized to operate in moderate inversion in order to optimise the figure of merit (FoM) ~\cite{Enz2015LowpowerAC}:
\begin{equation}
    \textrm{FoM} = \frac{g_\textrm{m}*f_\textrm{t}}{I_\textrm{d}}.
\end{equation}
This important parameter describes the behaviour of the transistor and it is defined as: $g_\textrm{m}/I_\textrm{d}$ that is the transconductance efficiency, while $f\textrm{t}$ is the transit frequency of the transistor. The FoM allows to find the size and the operating region of the transistor that maximize the bandwidth of the system for a fixed power consumption.

The first stage is followed by a differential amplifier with unitary gain feedback, needed to separate the amplification from the output load capacitance.
During the design phase, a source follower buffer was also considered. However, the voltage drop required to maintain the transistors in saturation mode drastically reduces the dynamic range, which is why the differential configuration was preferred.
The output of the buffer is then closed in a resistive feedback loop with the gate of the input transistors to stabilise the gain.
The amplifier and the buffer previously described are designed for a power consumption of 200~$\mu$W.

The peculiarity of the engineered monolithic device, as described in Section~\ref{subsec:sensor}, is the polarisation of the collection electrode to a voltage above 20~V. Therefore, it is crucial to pay special attention to the connection of the sensor to the front-end electronics as the latter operates at a much lower voltage (1.2~V). The proposed solution was to AC couple the pixel and the electronics.
The coupling capacitor is an interdigitated multi-finger capacitor implemented above the pixel area. Post-layout simulations were run to set the capacitance value to minimise signal attenuation, dimension, and parasitic to the ground while keeping the inter-finger distances sufficiently large to sustain the required voltage drop.

The front-end is then linked to an output buffer implemented as a source follower enclosed within two transistors working as switches (S1 and S2).
An inverter stage is used to turn the buffer on and off with only one signal.
Unlike the front-end, which was developed in 1.2~V, the 3.3~V MOSFETs were preferred for this stage since the transistors designed in this power domain have a wider dynamic range and a higher driving capability. 
Indeed, the transistors of this part are sized to drive a capacitive load higher than 20~pF with a power consumption of 1.6~mW per channel.

As for the collection electrode, having two different power domains implies that a direct connection is not possible. Thus, the output buffer is AC coupled (Cc1) to the front-end electronics and the polarization of the M1 gate is controlled by an external voltage through a 1~M$\Omega$ resistor (Rb1).

\subsection{Chip layout}
\begin{figure}[!ht]
\centering
     \begin{subfigure}[!ht]{\linewidth}
         \centering
         \includegraphics[width=0.8\textwidth]{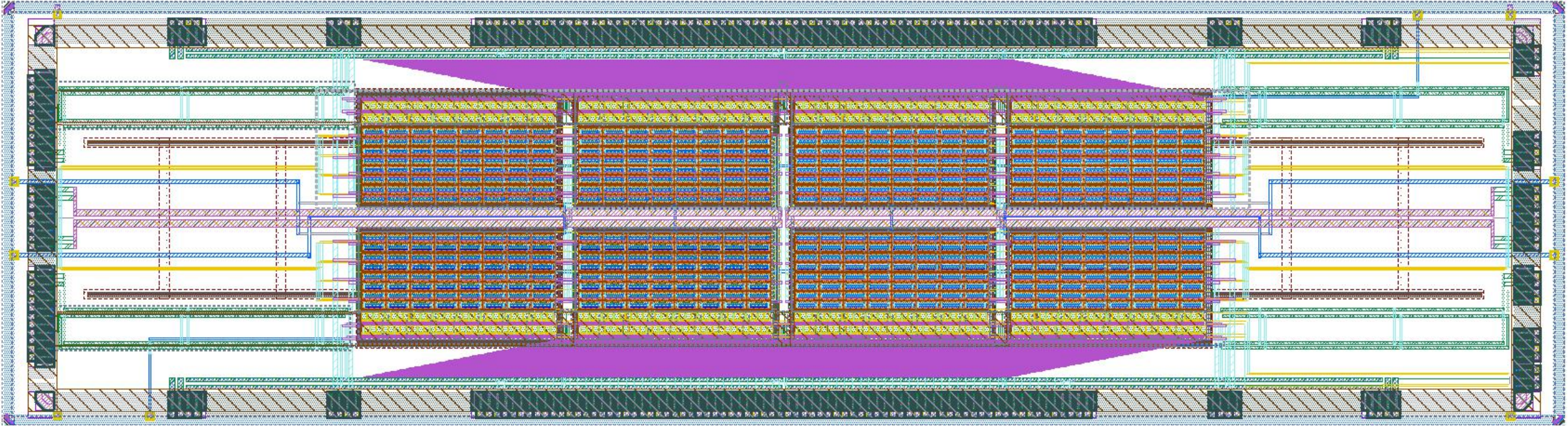}
     \end{subfigure}
     \vspace{4px} \break
     \begin{subfigure}[!ht]{\linewidth}
         \centering
         \includegraphics[angle=270,width=0.79\textwidth]{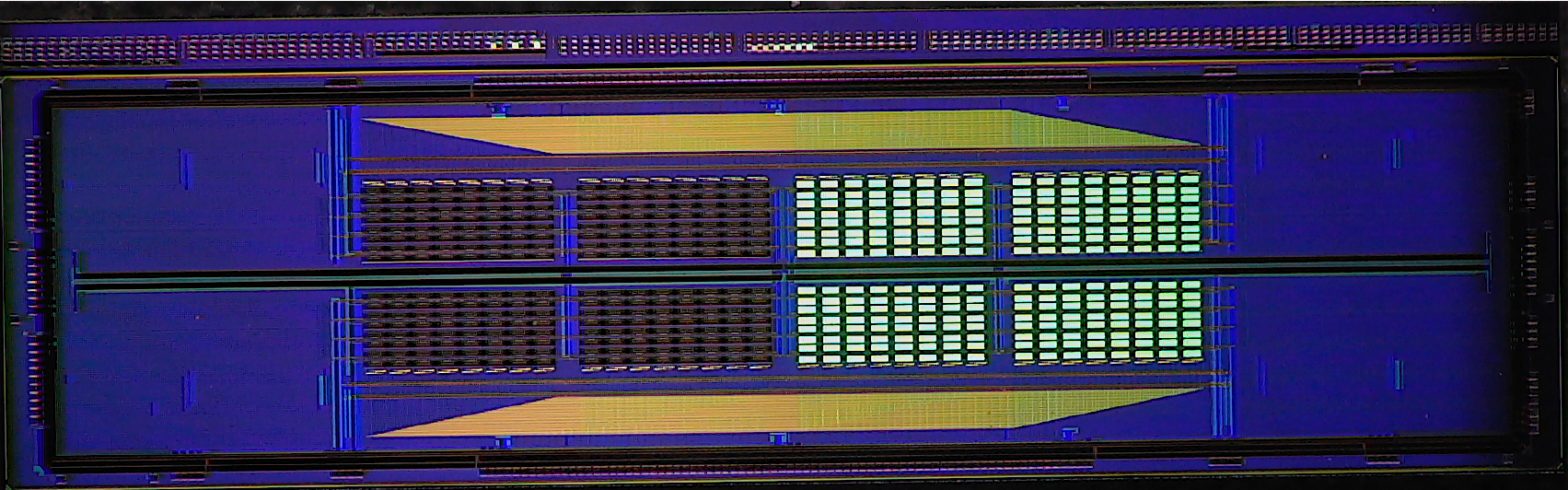}
     \end{subfigure}
\caption{Chip layout on the top and dice micrograph on the bottom.}
\label{fig:chiplayout}
\end{figure}

MadPix is 16.4~$\times$~4.4~mm$^2$ in size and includes two symmetrical adjacent regions: \textit{top} and \textit{bottom}.
Each one has four matrices composed by 64 pixels divided in 8 rows and 8 columns and the layout is shown in Figure~\ref{fig:chiplayout}.

Due to size constraint, the chip is designed with only 64 analogue outputs per each region, hence the Vout (Fig. \ref{fig:schematic}) of four 3.3~V buffers, one for each matrix, shares the same output pad, i.e. the output buffers are multiplexed in groups of four.
The pixel size is 100~$\times$~250~$\mu$m$^2$, while the 1.2~V front-end electronics is placed in a 8~$\times$~250~$\mu$m$^2$ \textit{p}-well implemented in the long side dividing two adjacent pixels.
Moreover, the 3.3~V source follower buffers are placed outside the matrices and inside the external guard-ring, an important element surrounding each matrix that, polarized to a positive voltage below 10~V, collects the electrons generated in the chip periphery, thus maintaining a low dark current in the pixels.
The ratio between the collecting electrode area and the total pixel area is around 67\%.
For timing measurement this ratio has to be as high as possible in order to minimise the distortion contribution to the time resolution~\cite{Riegler_2017}. However, the density rule of the process did not allow the use of larger pixels for this specific run. Still, the technology is compatible with bigger collection electrodes. 

The four matrices are labeled A1\_\textit{a}, A2\_\textit{a}, A1\_\textit{b} and A2\_\textit{b}. The A1 and A2 labels distinguish the two layout flavour described in Section~\ref{sec:design}. Instead, the letters \textit{a} or \textit{b} represent two different layouts of the coupling capacitor. Indeed, the library element supplied by the foundry is guaranteed to work up to 30~V while the top electrode needs to be polarized up to 50~V to achieve the target value of the gain. Therefore, a full custom capacitor (layout \textit{b}) was designed with a larger gap between the plates to avoid breakdown discharge. Only the A1\_\textit{a} and A2\_\textit{a} flavors were characterised because no breakdown was observed with the standard capacitor up to 50~V.

As for the readout, two different configurations are possible: a single matrix or a quadruplet pixel readout.
The former is active if a matrix $Enable$ signal is high, then all the pixels of a matrix are linked to the 64 analogue outputs, and the lines can be read in parallel.
The latter enters into operation when all the $Enable$ are low; with this configuration only four adjacent pixels of a matrix are selected through a shift-register and read out. The remaining pixels are switched off.

\subsection{Readout System}
\begin{figure}[!ht]
\centering
     \begin{subfigure}[!ht]{0.31\linewidth}
         \centering
         \includegraphics[width=\textwidth]{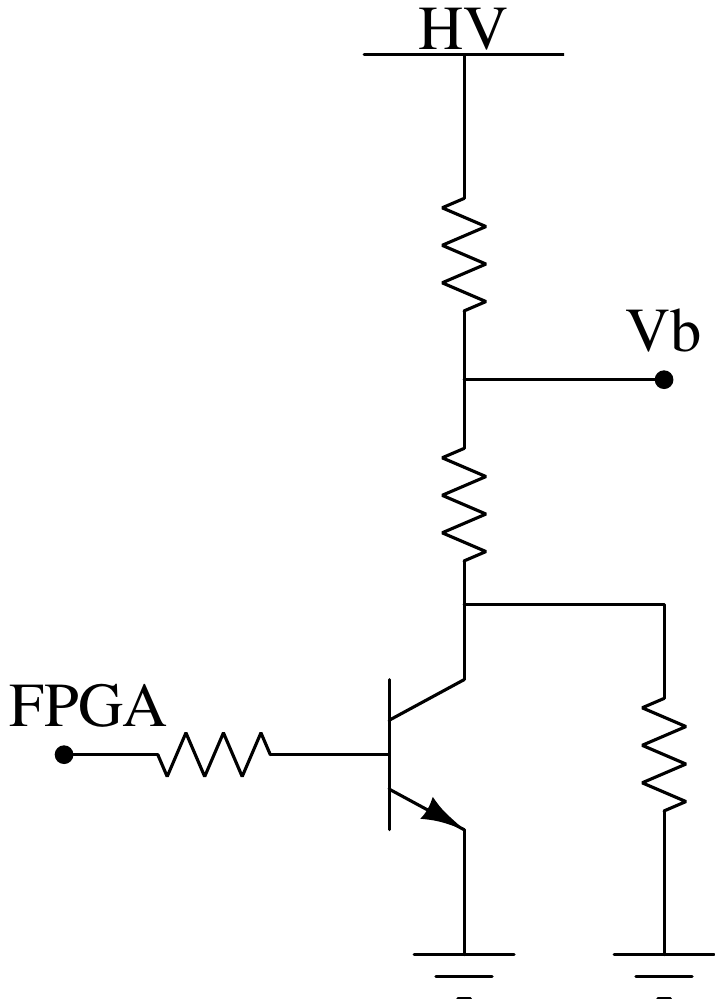}
         \caption{}
         \label{fig:TP_schematic}
     \end{subfigure}
     \hfill
     \begin{subfigure}[!ht]{0.49\linewidth}
         \centering
         \includegraphics[angle=180,width=\textwidth]{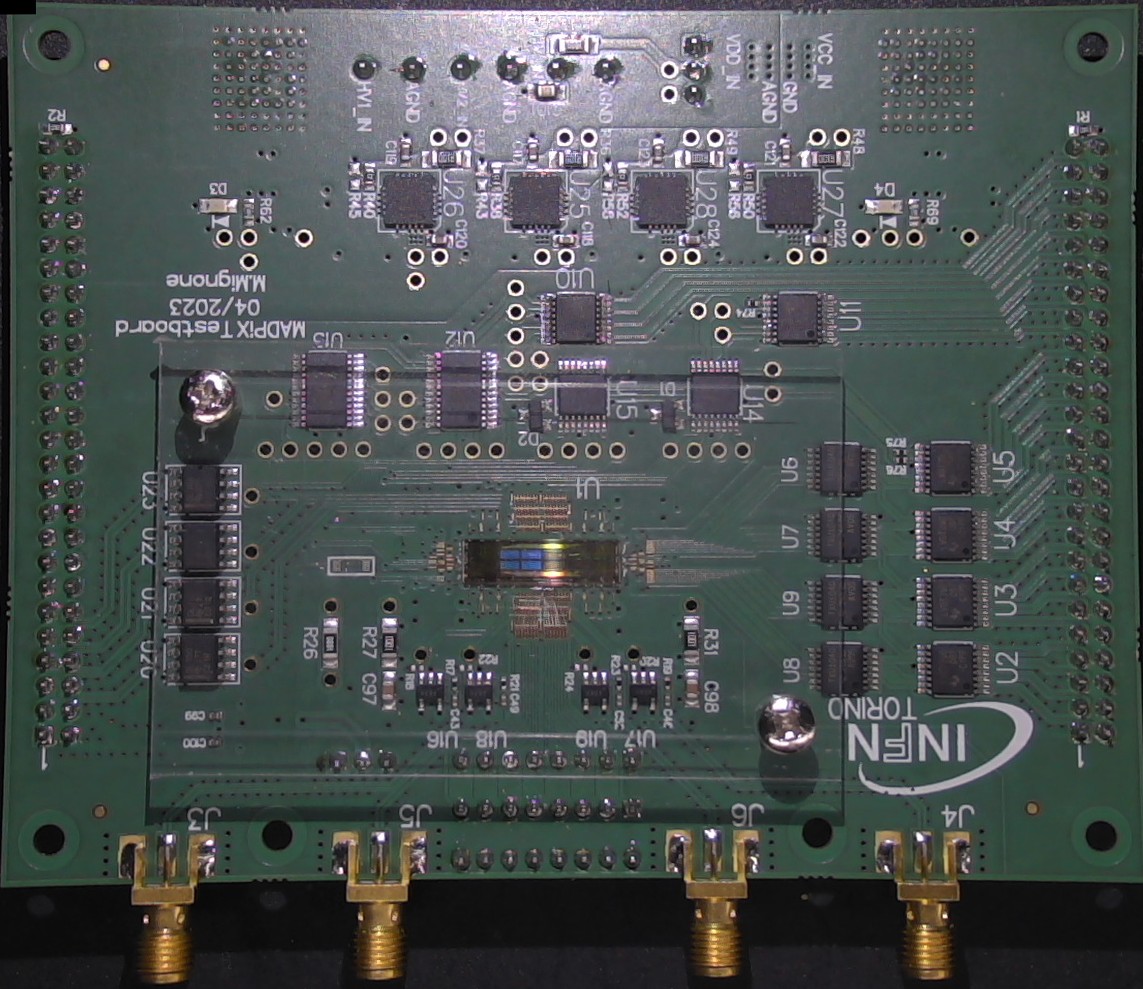}
         \caption{}
         \label{fig:testboard}
\end{subfigure}
\caption{Schematic of the test pulse circuit placed on the test board (a) and picture of the PCB (b). The ASIC is wire bonded to the PCB.} 
\end{figure}

A custom PCB TestBoard (TB) has been developed in order to test the MadPix chip (Figure~\ref{fig:testboard}).
Its main purpose is to set all the bias currents and voltages needed by the electronics.
In addition, the board provides all the digital signals employed to select the readout mode.

Like the chip, the PCB is divided in two parts: \textit{top} and \textit{bottom}.
The top has four analogue buffers each one driving a 50~$\Omega$ impedance line, that can be connected to an oscilloscope or to a digitizer via a SMA cable.
The bottom part has four LVDS buffers used as discriminators, with the outputs sent to an external FPGA.
The same FPGA controls the board via two flat cables (IC 50 connectors placed on the back of the TB, not visible in the picture).

Depending on the wire bonding configuration, the four analogue buffers and the four discriminators placed on the TB can alternatively read all the pixels or just some of them.
Indeed, 16 of the 64 top and bottom outputs pads can be bonded in parallel on the test board in order to read a quadruplet of pixel. The selection of the quadruplet is made by a shift register and all the remaining pixels are switched off thanks to S1 and S2 in Figure~\ref{fig:schematic}.

The PCB is also responsible for generating the test pulse signal that can be used to easily check the operation properties of the chip.
The signal is generated using the circuit displayed in Figure~\ref{fig:TP_schematic}: the BJT is turned on with a signal coming from the FPGA and, as a result, a step in the HV is generated. This perturbation is then propagated through all the chip as the HV line is common to all pixels.

\section{Test bench results}
This section presents the results obtained from the electrical and optical characterisations.

\subsection{Sensor characterisation}
\begin{figure}[!ht]
\centering
     \begin{subfigure}[!ht]{0.49\linewidth}
         \centering
             \includegraphics[width=\textwidth]{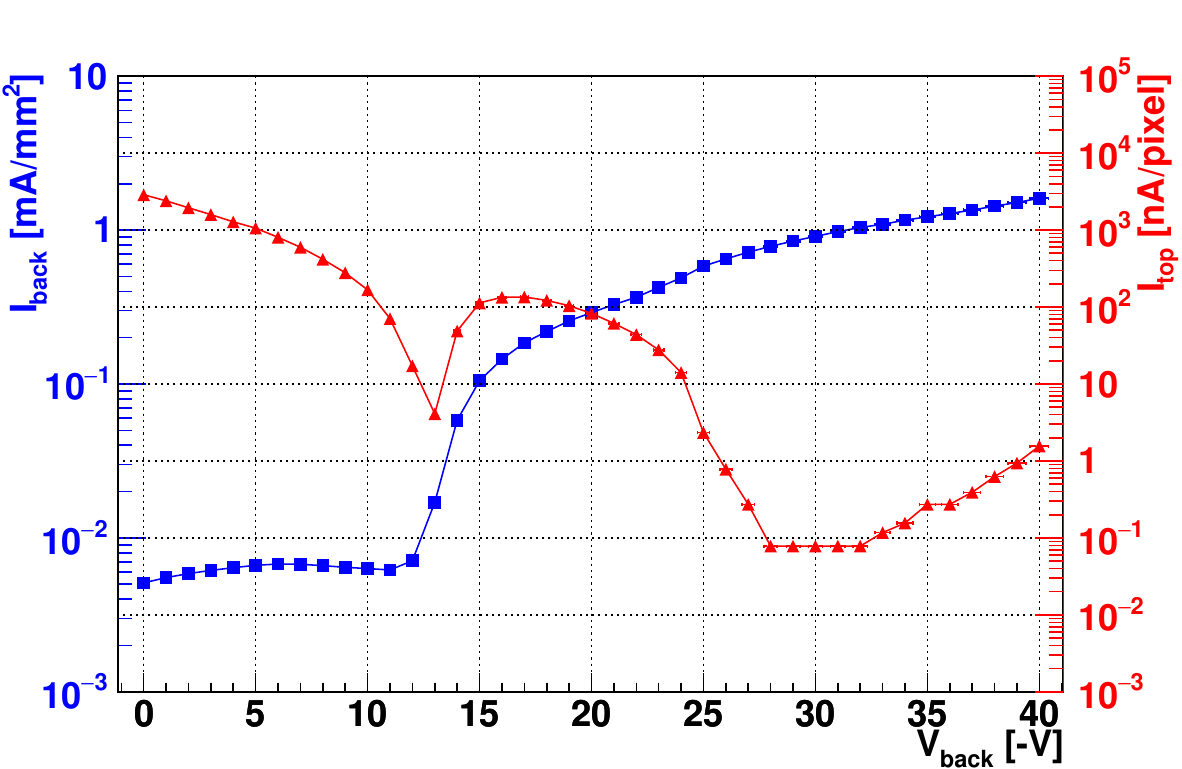}
         \caption{}
         \label{fig:ivback}
     \end{subfigure}
     \hfill
     \begin{subfigure}[!ht]{0.49\linewidth}
         \centering
         \includegraphics[width=\textwidth]{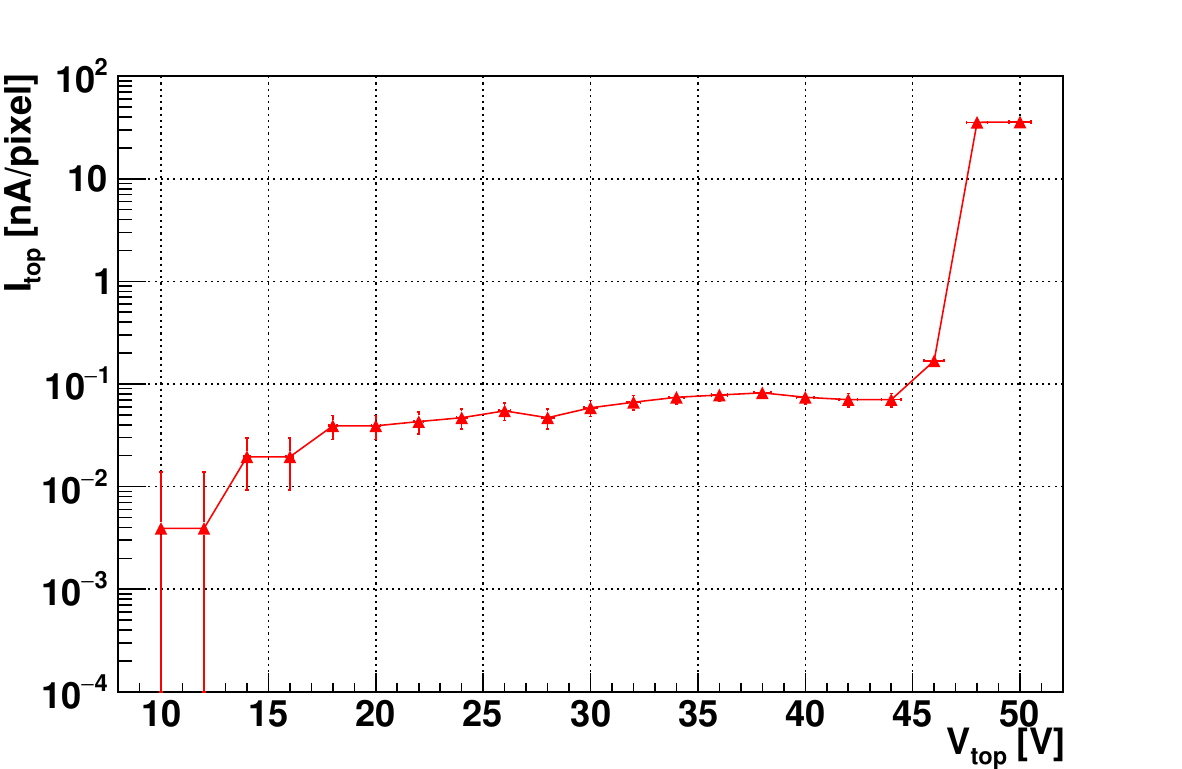}
         \caption{}
         \label{fig:ivtop}
\end{subfigure}
\caption{IV plots of the backside (blue squares) and topside (red triangles) currents as a function of the backside (a) and of the topside (b) voltage. The plot (a) is obtained with a given voltage on the collection electrode of 45~V, while plot (b) is reported for $-30$~V applied to the backside.}
\label{fig:iv}
\end{figure}

Current versus voltage $I(V)$ measurements were performed to characterise the sensor. Specifically, two currents were measured as a function of the scan of the two polarisation voltages with the external guard-ring (\textit{n}-type implant) biased at 9~V (which is high enough to facilitate the fully depletion of the sensor, but not too high avoiding an edge breakdown): the backside current $I_\textrm{back}$ and the topside current $I_\textrm{top}$.  
The first one is the current flowing from the \textit{p}-doped electrode located on the backside of the sensor and it strongly increases due to the punch-through effect which usually arises at backside voltages above 35~V. The $I_\textrm{top}$ is the current flowing at the collection electrodes and it significantly grows due to breakdown phenomena when the top electrode is polarised above 45~V.
The results are reported in Figure~\ref{fig:iv} where the backside current is normalised per mm$^{2}$ while the topside current is given per pixel.

The plot~\ref{fig:ivback} illustrates the currents ($I_\textrm{back}$ and $I_\textrm{top}$) as a function of the backside voltage $V_\textrm{back}$ for the case of 45~V applied on the top collection electrode $V_\textrm{top}$. The $I_\textrm{top}$ current shows a decreasing trend with increasing reverse bias, due to the progressive depletion process. Indeed, at low $V_\textrm{back}$ the $I_\textrm{top}$ is composed mainly by a flow of charge from the collection electrode to the \textit{n} guard-ring placed around the pixels. Once the substrate is depleted and the conductive path between these two structures is eliminated, we can easily appreciate the thermally generated carriers entering the gain region and causing an increase of the top current. As can be observed, the backside voltage required to fully deplete the substrate is around $-28$~V, where the $I_\textrm{top}$ stops decreasing and reaches a range of stability. Furthermore, the backside current has a significant increase around $-12$~V due to a floating guard-ring placed outside the padframe, i.e. very close to the silicon border where the chip is diced. This element, if biased, would collect the increase of the border current due to the low resistive path created during the dicing process. This growth has also an impact on the $I_\textrm{top}$, which rises when the backside voltage exceeds $-12$~V (see Figure~\ref{fig:ivback}). Indeed, the high $I_\textrm{back}$ shifts the start of the full depletion to $-28$~V.
Moreover, it is possible to observe that for reverse bias higher than $33$~V, the topside current starts to increase as the multiplication process is increasing too. In Figure~\ref{fig:ivtop} a scan of the $V_\textrm{top}$ was made polarizing the $V_\textrm{top}$ at -30~V, and the breakdown voltage can be estimated to be around 47~V when the current starts increasing exponentially. 

\subsection{Test pulse characterisation}
This section illustrates the characterisation of the pixels of the two top matrices using the test pulse circuitry. The analogue amplitudes in response to an injected test pulse were measured using a Lecroy WaveSurfer 4104HD oscilloscope. 

\begin{figure}[!ht]
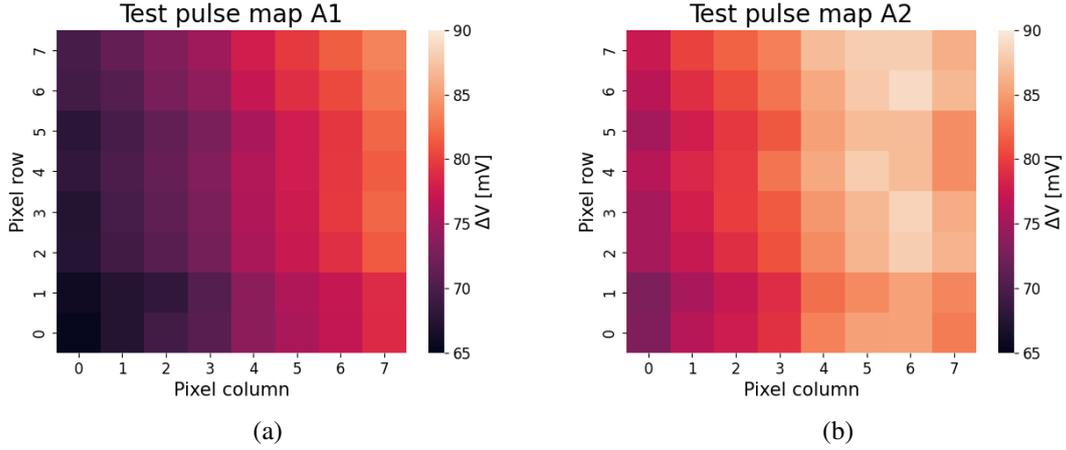

\centering
     \begin{subfigure}[!ht]{0.49\linewidth}
         \centering
         \includegraphics[width=\textwidth]{TP_A1}
         \caption{}
         \label{fig:TP_A1}
     \end{subfigure}
     \begin{subfigure}[!ht]{0.49\linewidth}
         \centering
         \includegraphics[width=\textwidth]{TP_A2}
         \caption{}
         \label{fig:TP_A2}
     \end{subfigure}
\caption{Analogue amplitude measured for each pixel of matrices with flavour A1 (a) and A2 (b).}
\label{fig:TP}
\end{figure}

The results obtained for the two layouts described in Section~\ref{subsec:sensor} (A1\_\textit{a} and A2\_\textit{a}) are reported in Figure~\ref{fig:TP}. A pattern in the pixels response is evident along the matrix due to the layout of the power rails, which are routed from the right side to the left side for both A1 (Figure~\ref{fig:TP_A1}) and A2 (Figure~\ref{fig:TP_A2}) arrays. Indeed, the signal amplitude decreases with the decrease of the column number due to the voltage drop. A 20\% variation on the output amplitude was observed overall in each matrix and no dead pixels were found.

\subsection{Laser measurement}

After the electrical characterisation, the behaviour of the chip has been investigated using the Transient Current Technique (TCT), which exploits the signal induced in the sensor by a laser. The TCT setup employed in this study is equipped with micro-metrical \textit{x-y} step motors and a fast pulsed (100~ps) IR laser (1060~nm wavelength), that can be focused in a minimum spot with a FWHM $\sim$10~$\mu$m.

Four analogue outputs of MadPix have been wire-bonded to the 4-channel readout board. The output signals have been aqcuired with a digital oscilloscope having 1~GHz bandwidth (Tektronix MDO3102). 

\begin{figure}[!ht]
\centering
     \begin{subfigure}[!ht]{0.49\linewidth}
         \centering
         \includegraphics[width=\textwidth]{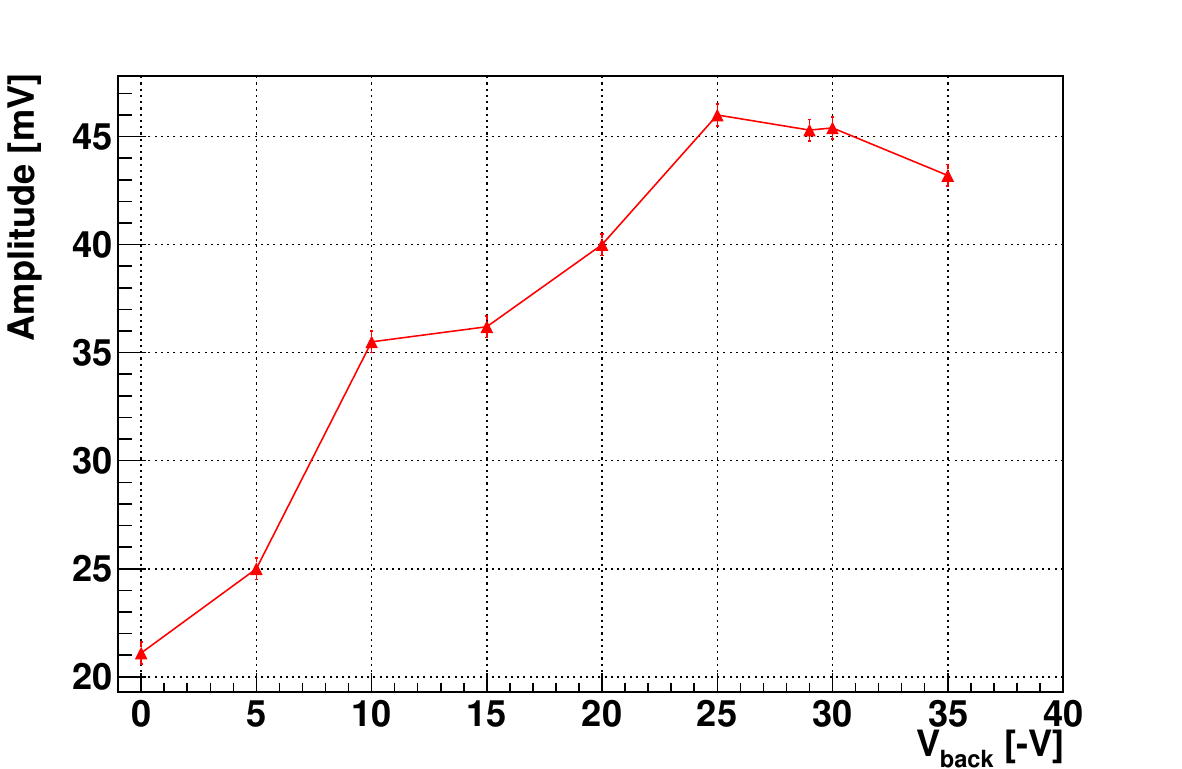}
         \caption{}
         \label{fig:laserdepletion}
     \end{subfigure}
     \begin{subfigure}[!ht]{0.49\linewidth}
         \centering
         \includegraphics[width=\textwidth]{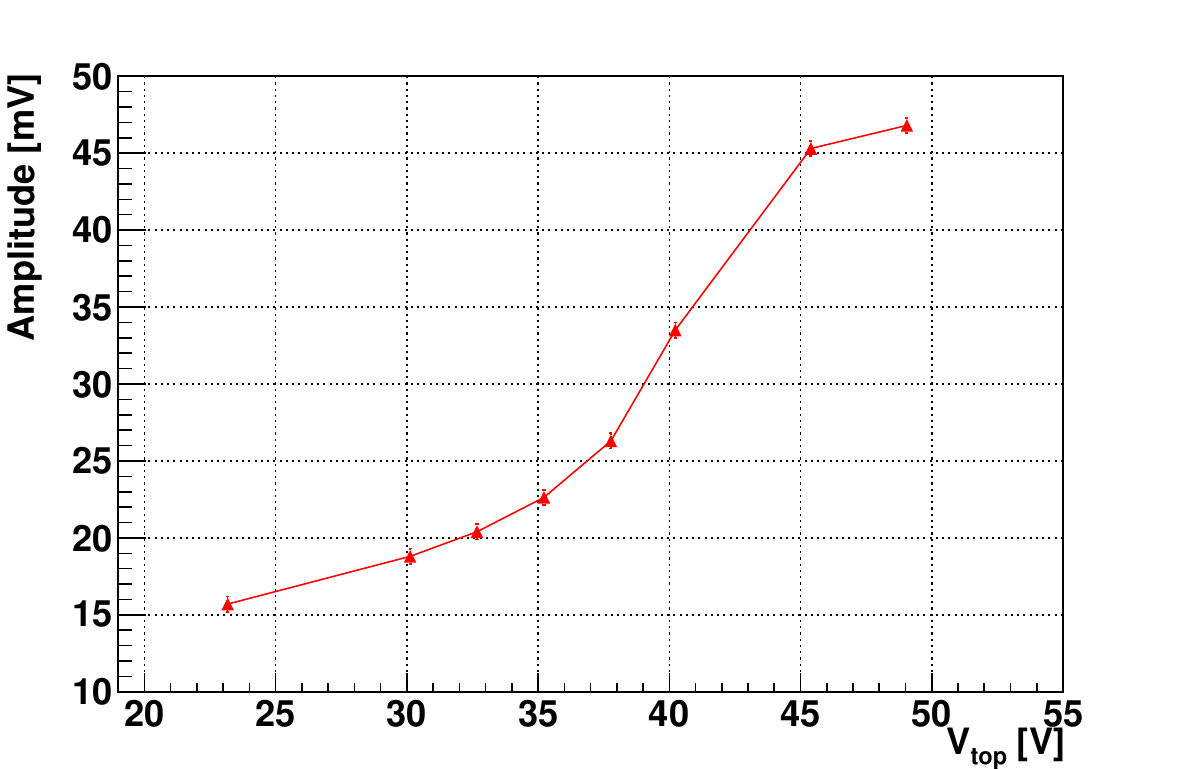}
         \caption{}
         \label{fig:lasergain}
     \end{subfigure}
\caption{Analogue output amplitude of one pixel of flavour A1 in response to a laser pulse. The measurements were carried out changing the backside voltage (a) and the collection electrode bias voltage (b). The V$_\textrm{top}$ was set at 45~V in (a) while in (b) the V$_\textrm{back}$ to $-30$~V.}
\label{fig:laser}
\end{figure}

First, the value of the depletion voltage and the bias dependence of gain have been investigated. 
The analogue output signal of a pixel in response to a laser pulse of constant intensity has been acquired by varying the top and back voltages. The signal amplitude as a function of both V$_\textrm{top}$ and V$_\textrm{back}$ is illustrated in Figure~\ref{fig:laser}. Specifically, the depletion voltage, i.e. voltage at which both the substrate and gain layer (since V$_\textrm{top}$ is at 45~V) are fully depleted and the drift velocity of the carriers is saturated, corresponds to the point where the amplitude of the signal stops increasing while decreasing the backside voltage, i.e. V$_\textrm{back}<25$~V (Fig.~\ref{fig:laserdepletion}). On the other hand, from Figure~\ref{fig:lasergain}, it can be observed that rising the voltage applied to the frontside collection electrode from 30~V to 45~V, the signal is enhanced by a factor $\sim$~2.5, proving that the gain gradually grows in this range of V$_\textrm{top}$. The obtained value is in agreement with the one obtained from simulations (Sec. \ref{subsec:gainprofile}).

The same setup has been exploited to provide an estimation of the electronic jitter. 
The most important contributions to the temporal resolution $\sigma_\textrm{t}$ are shown in equation~\ref{eq:time resolution}:
\begin{equation}
\label{eq:time resolution}
\begin{aligned}
\sigma_\textrm{t}^{2} \simeq \sigma_\textrm{Time Walk}^{2} + \sigma_\textrm{Landau Noise}^{2} + \sigma_\textrm{Distortion}^{2} + \sigma_\textrm{Jitter}^{2} + \sigma_\textrm{TDC}^{2}
\end{aligned}
\end{equation}
Each of these terms takes into account a different physical effect and influences the time resolution. A detailed description can be found in~\cite{2021Ferrero_CRC}.

For the specific case of a laser pulse, the time resolution is not affected by Landau and distortion contributions (Eq.~\ref{eq:time resolution}), if the laser spot is smaller than the pixel size. For this measurement, it is important to increase by steps the IR laser intensity performing a laser power scan since the jitter is function of the signal amplitude. In this way, the time walk contribution is negligible as the intensity of the laser is fixed at each step. 

\begin{figure}[!ht]
\centerline{\includegraphics[width=0.70\linewidth]{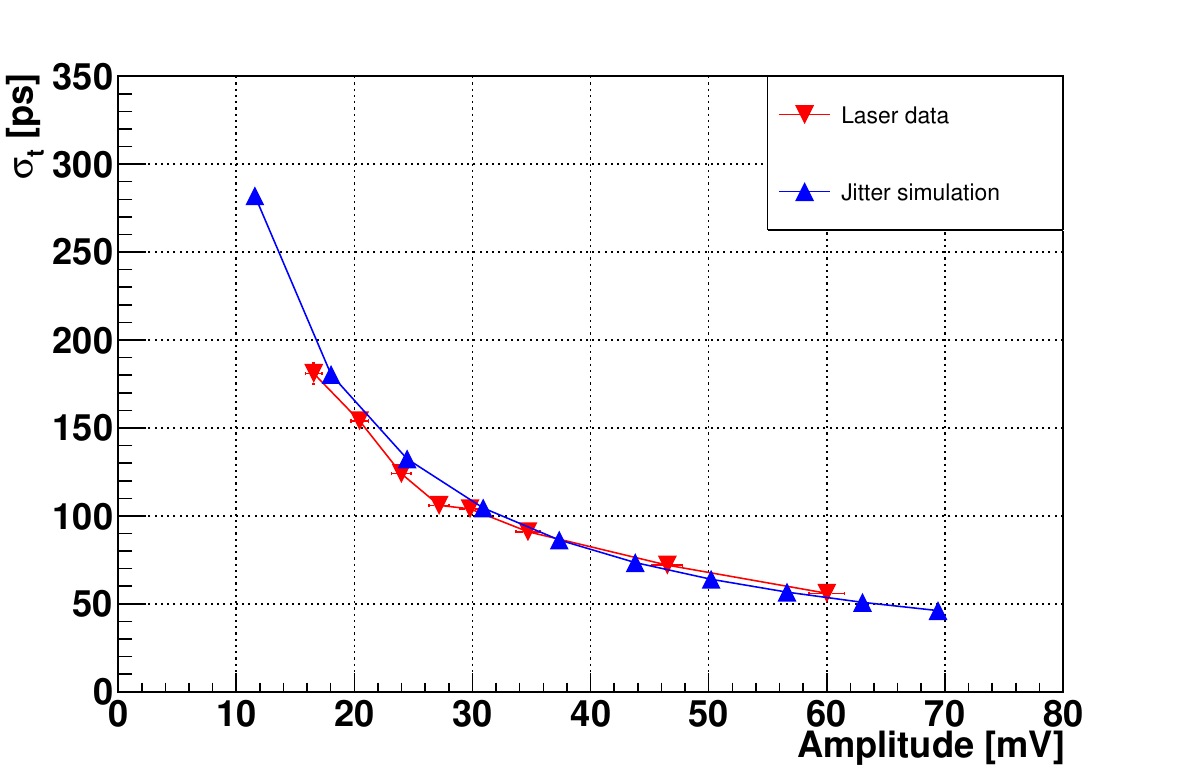}}
\caption{Electronic jitter extracted from circuit simulations (blue) and time resolution from experimental laser measurements (red), acquired with V$_{back}$~=~-30~V and V$_{top}$~=~40~V.}
\label{fig:jitterlaser}
\end{figure}

Initially, the laser has been set up to release an amount of charge similar to the one induced by one Minimum Ionizing Particle (MIP) and the time resolution was estimated as the standard deviation of the time difference between laser trigger out (TTL) and the analogue output of MadPix at 50$\%$ signal amplitude.
The measurement results are reported in Figure~\ref{fig:jitterlaser} and, in the same plot, the jitter evaluated in post-layout simulations of the electronics is shown. Specifically, the capacitance of the detector used in these simulations was extracted using the CV of the passive structures (Fig.~\ref{fig:CV}) and the jitter was evaluated using the definition~\ref{eq:jitter}:
\begin{equation}
    \label{eq:jitter}
    \begin{aligned}
    \sigma_\textrm{jitter} \propto \frac{N}{\frac{\textrm{d}V}{\textrm{d}t}},
    \end{aligned}
\end{equation}
where the slew rate $\textrm{d}V/\textrm{d}t$ is measured at \mbox{40-60\%} of the signal amplitude and the RMS noise is estimated from the integral of the power spectral density obtained through an AC noise analysis. 

As expected, by increasing the signal amplitude, raising the output power of the laser, the jitter contribution decreases achieving $\sim50$~ps.

\section{Test Beam measurements}
In the following section the test beam setup is described. The charge distributions and the time resolution are then analysed.

\subsection{Setup}
The last step to complete the characterisation of MadPix is the measurement of the time resolution. The test beam campaign was carried out at the Proton Synchrotron at CERN (T10 beamline) in October 2023, where a beam composed by a mixture of proton and pions with a momentum of 10~GeV/c was provided. 
The sensors were arranged on moving stages with micrometric precision and the telescope was placed in a dark box kept at room temperature.
The telescope was made of two planes: the Device-Under-Test (DUT) and a 1~$\times$~1~mm$^2$ LGAD manufactured by FBK~\cite{2020Sola_RD50} with a thickness of 35~$\mu$m, employed as trigger plane.
The outputs of three adjacent pixels of MadPix were read out and acquired together with the LGAD output. In particular, a LeCroy WaveRunner \mbox{9404M-MS} oscilloscope with 20~GS/s sampling rate, 4~GHz of analogue bandwidth and \mbox{8-bit} vertical resolution was used. Since only four signals could be recorded simultaneously, only the data of three adjacent pixels were acquired and the last channel of the oscilloscope was used for the reference trigger. Specifically, the data acquisition was performed triggering on the coincidence of one pixel of MadPix and the LGAD.

\subsection{Collected charge}
\begin{figure}[!ht]
\centering
     \begin{subfigure}[!ht]{0.49\linewidth}
         \centering
         \includegraphics[width=\textwidth]{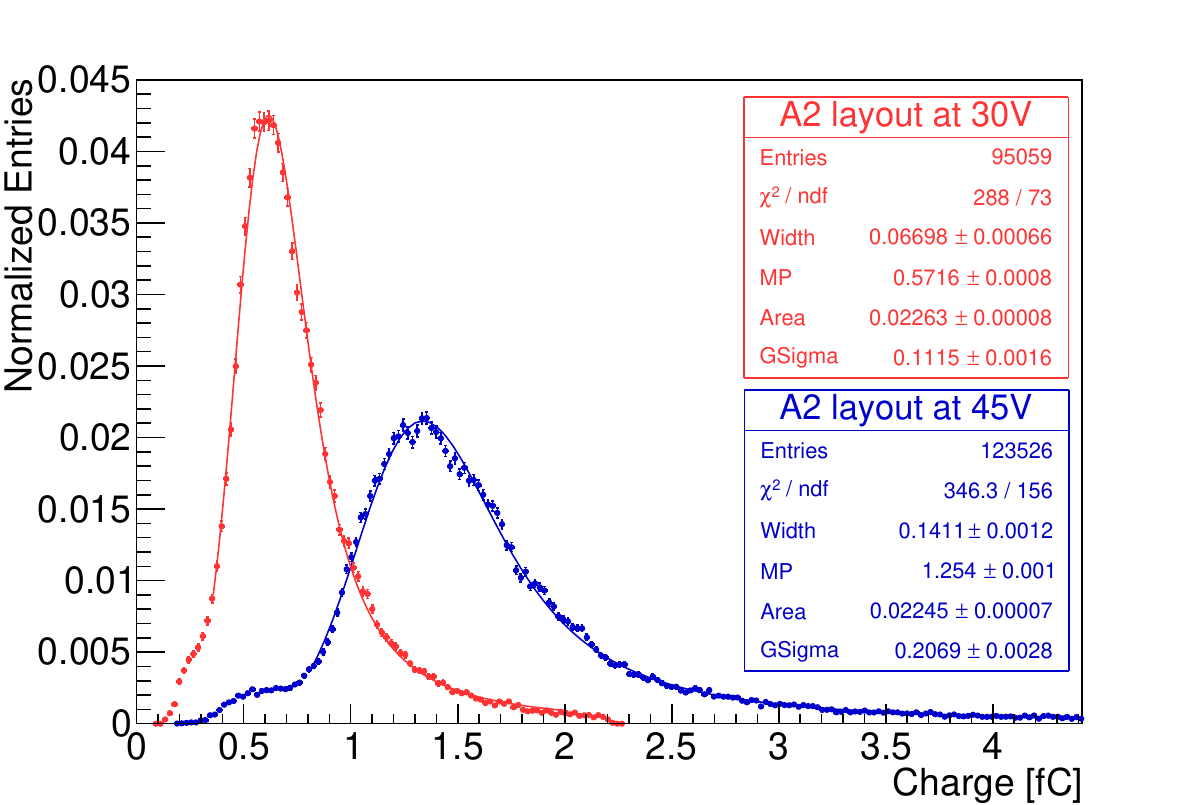}
         \caption{}
         \label{fig:ChargeDistribution_V_different}
     \end{subfigure}
     \begin{subfigure}[!ht]{0.49\linewidth}
         \centering
         \includegraphics[width=\textwidth]{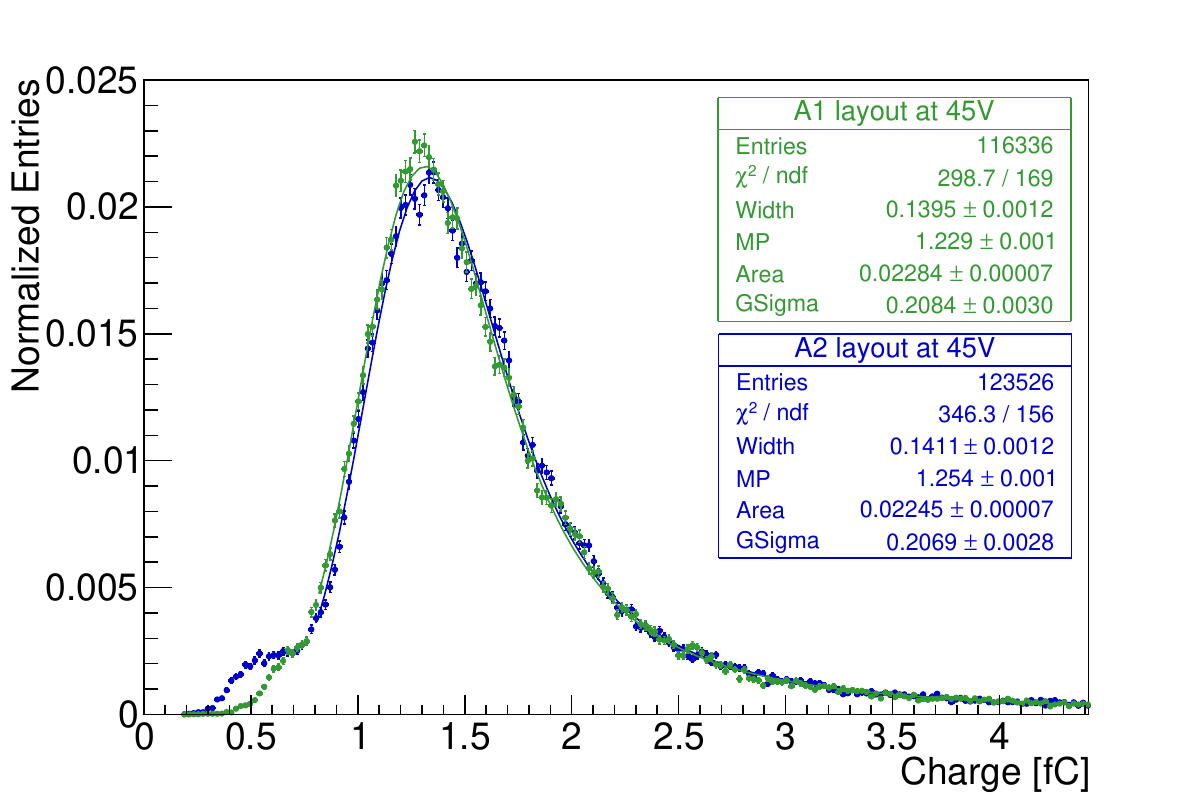}
         \caption{}
         \label{fig:ChargeDistribution_A1andA2}
     \end{subfigure}
\caption{Distribution of the collected charge for two different top bias voltages in the A2 structure: 30~V in red and 45~V in blue (a). Distribution of the charge for the two structures (V$_\textrm{top}$~=~45~V): A1 in green and A2 in blue (b). The backside polarization was set at $-30$~V.}
\label{fig:ChargeDistribution}
\end{figure}

Owing to the sampling of the full analogue waveform, the collected charge distribution could be extrapolated and the results are reported in Figure~\ref{fig:ChargeDistribution}.
In these plots, the charge is evaluated as the maximum amplitude of the analogue signal divided by the electronics gain extracted through post-layout simulations and the distributions are fitted with a convolution of a Landau and a Gaussian functions. 
In Figure~\ref{fig:ChargeDistribution_V_different} the charge distributions of the A2 structure are plotted for different collection electrode voltages, i.e. 30~V and 45~V. As can be observed, by increasing the voltage, the MPV of the distribution increases by a factor $\sim2.2$, which confirms the activation of the gain, in agreement with the laser measurements.
Figure~\ref{fig:ChargeDistribution_A1andA2} illustrates the comparison between the structures A1 and A2 at the same bias voltage on the collection electrode, which was set at 45~V. The small difference in the MPVs could be explained by minor fluctuations in the gain of the electronics, while the mismatch in the tails at small charge values can be attributed to the different termination of A1 and A2, since the former provides charge gain through avalanche multiplication also at the edges of the sensor.

\subsection{Time Resolution}
\label{subsec:Time Resolution data analysis}

The full waveforms of the LGAD trigger plane and the DUT have been recorded and analysed. The time resolution of Madpix has been evaluated with an offline analysis software and exploiting different analysis methods (fixed threshold and time at a constant fraction). This paper reports the best results obtained using a fixed threshold algorithm with time walk correction.

\begin{figure}[!ht]
\centerline{\includegraphics[width=0.70\linewidth]{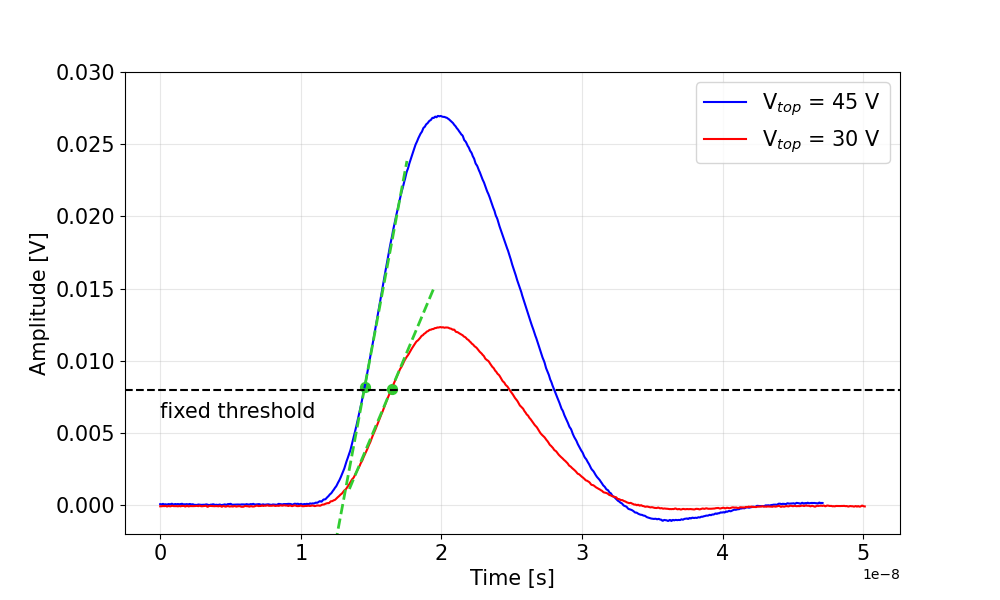}}
\caption{Averaged waveforms of 300 collected events acquired at a top voltage of 45~V and 30~V. A fixed threshold of 8~mV is shown as example, together with the tangent in correspondence of the crossing time. The mean amplitudes of the two waveforms are 27$\pm$10 mV and 13$\pm$5 mV, where the errors have been evaluated as the standard deviation of the distributions obtained with the aforementioned statistics.}
\label{fig:waveform}
\end{figure}

In this method, the Time Of Arrival (TOA) of a particle is set as the time where the signal crosses a given threshold in mV, as depicted in Figure~\ref{fig:waveform}. The plot shows two averaged waveforms obtained with two different top voltages: 45~V and 30~V.  
As an example, a constant threshold of 8~mV is applied. The lowest threshold applied has been chosen in order to be at least equal to 5~$\sigma_\textrm{noise}$.\\
For what concerns the trigger plane, the TOA has been evaluated using the crossing time at a 30\% fraction of the amplitude, obtained with the algorithm described in~\cite{2023Carnesecchi_EPJP}.

\begin{figure}[!ht]
\centering
     \begin{subfigure}[!ht]{0.51\linewidth}
         \centering
         \centerline{\includegraphics[width=\linewidth]{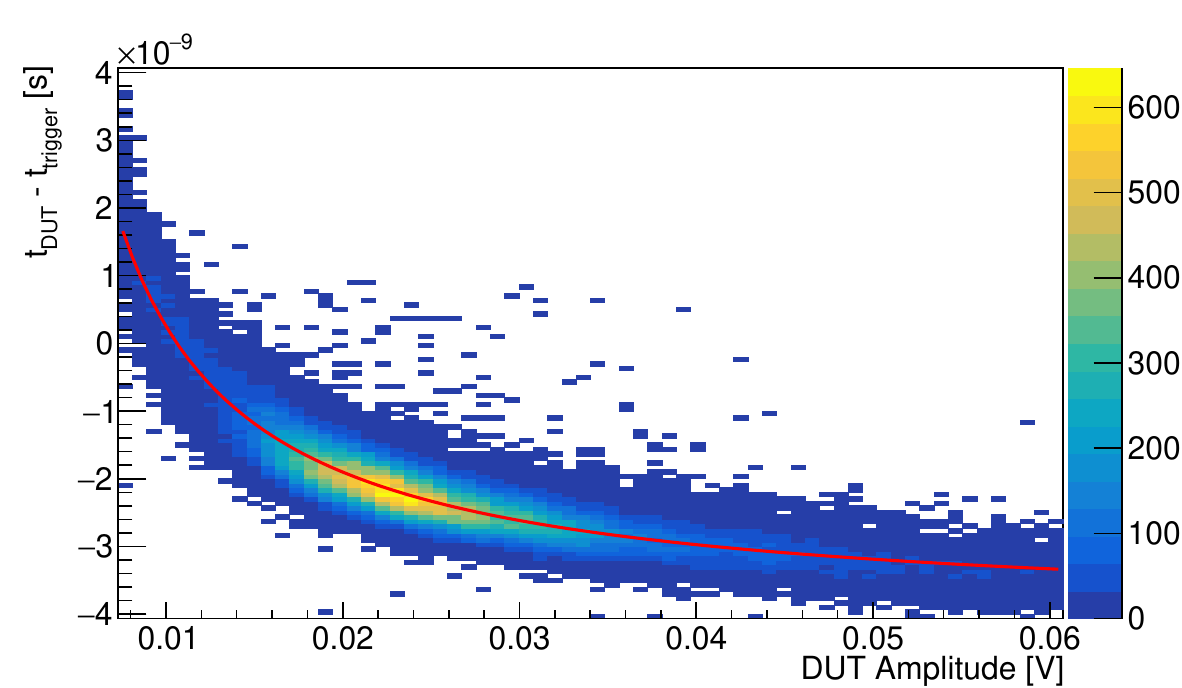}}
         \caption{}
         \label{fig:TimeDistributionVsAmpl}
     \end{subfigure}
     \begin{subfigure}[!ht]{0.48\linewidth}
         \centering
         \includegraphics[width=\textwidth]{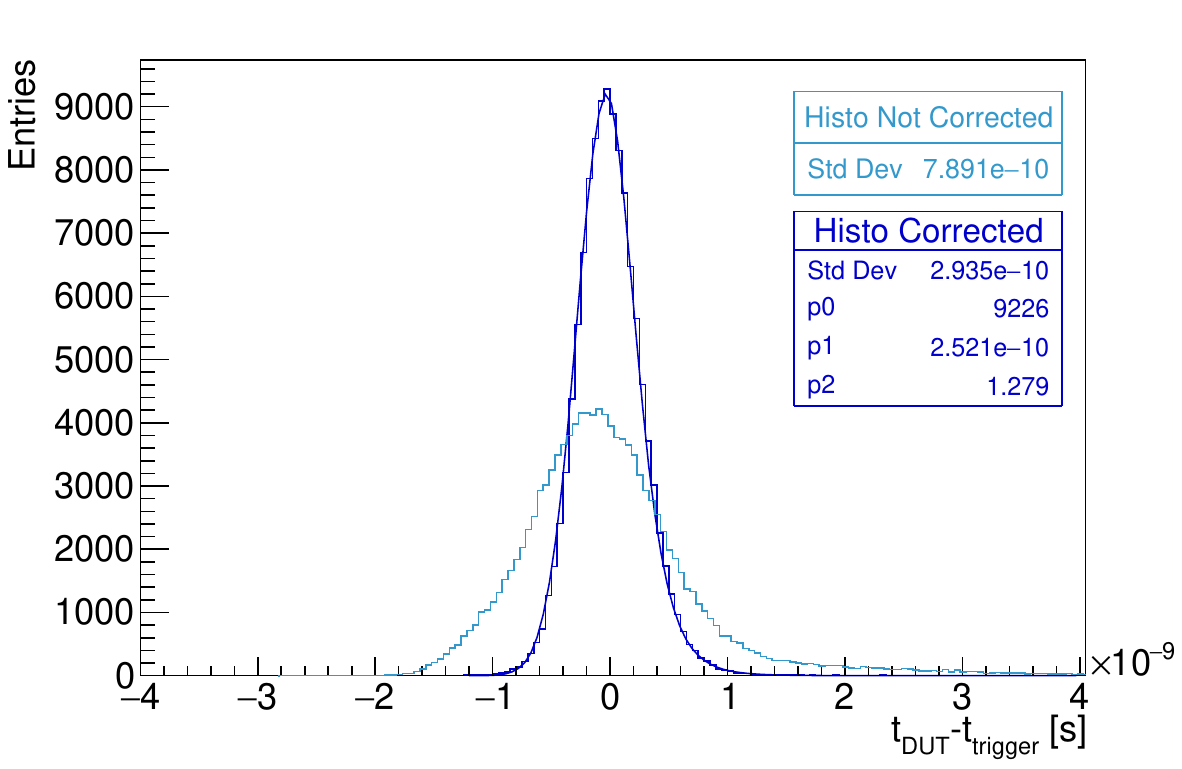}
         \caption{}
         \label{fig:TimeDifferenceLE}
     \end{subfigure}
\caption{Distribution of the time difference between the CMOS sensor and the trigger system as a function of the DUT amplitude. The red line is the fit used to perform the time-walk correction (a). Time difference distributions between the DUT and LGAD before
(light blue) and after (dark blue) the time-walk correction (b). The fit used to extract the time resolution is superimposed to the second histogram, where p0 represents the normalization, p1 the sigma, and p2 the right heavy tail of the q-Gaussian.
}
\label{fig:TimeDistribution}
\end{figure}

However, by employing the fixed threshold method, it has been necessary to correct the time-walk effect. Thus, the time differences between the DUT and the LGAD have been plotted as a function of the signal amplitude of DUT, as shown in Figure~\ref{fig:TimeDistributionVsAmpl}. The distribution is then fitted with the following function:
\begin{equation}
    t_{DUT} - t_{trigger} = a + \frac{b}{(DUT Amplitude)} + \frac{c}{(DUT Amplitude)^2}
\end{equation}
and the correction is applied by subtracting its value from the
measured data.
Figure~\ref{fig:TimeDifferenceLE} shows the time difference distribution obtained with a threshold of 5~mV applied on DUT, before and after the time walk correction.
Finally, the corrected time difference distributions were used to evaluate the time resolution of the sensor. The distributions are fitted with an asymmetric \mbox{q-Gaussian} function to take into account the tails. Afterwards, the resolution of the reference LGAD is quadratically subtracted from the sigma of the fits.

\begin{figure}[!ht]
\centering
     \begin{subfigure}[!ht]{0.49\linewidth}
         \centering
         \includegraphics[width=\textwidth]{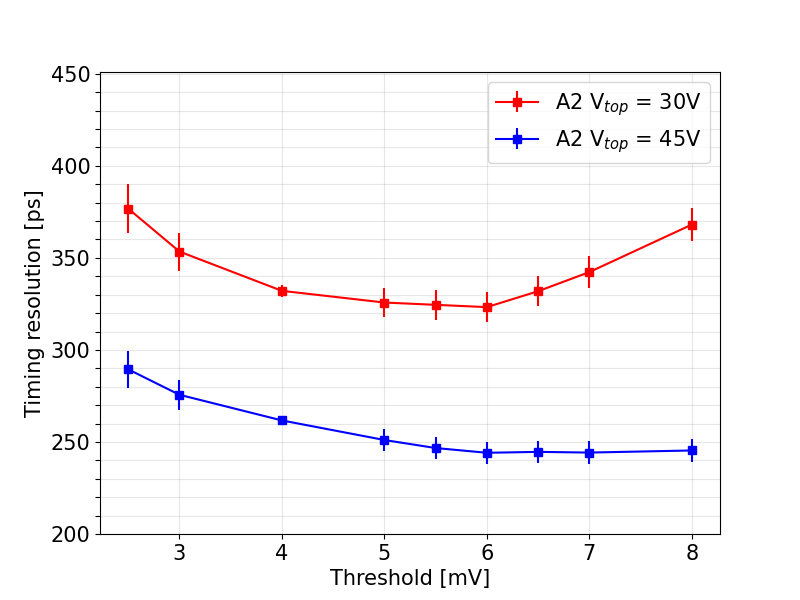}
         \caption{}
         \label{fig:A2_Vtop_30_45}
     \end{subfigure}
     \begin{subfigure}[!ht]{0.49\linewidth}
         \centering
         \includegraphics[width=\textwidth]{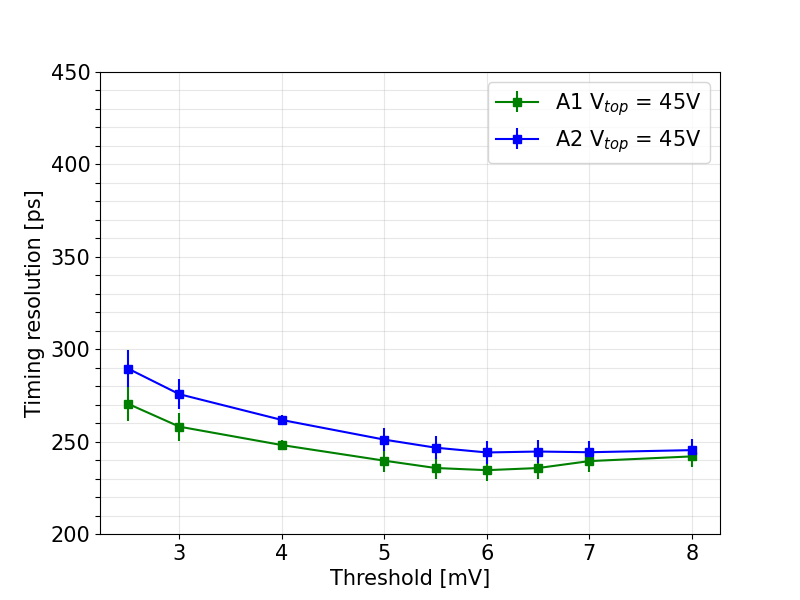}
         \caption{}
         \label{fig:Vtop45_A1_vs_A2}
     \end{subfigure}
\caption{Time resolution as a function of the applied threshold for an A2 structure at two different $V_\textrm{top}$ voltages (a). Time resolution at $V_\textrm{top}=45$~V for the two structures A1 and A2 (b). The $V_\textrm{back}$ is set at $-30$~V.}
\label{fig:timing_res}
\end{figure}

The time resolution obtained with the fixed threshold method is shown in Figure~\ref{fig:timing_res} as a function of different applied thresholds. Figure~\ref{fig:A2_Vtop_30_45} clearly illustrates the difference in the response of an A2 test structure when the top voltage is raised from 30~V to 45~V. On average, an improvement around 25$\%$ in the timing performance is observed with the $V_\textrm{top}=45$~V bias voltage. In the $V_\textrm{top}=30$~V condition, the avalanche gain is strongly reduced and this has an effect on the signal amplitude, that will be lower with respect to the $V_\textrm{top}=45$~V case. For this reason, the resolution at $V_\textrm{top}=30$~V starts to increase for thresholds higher than 6~mV, where the slope of the signal starts to decrease, as shown in Figure~\ref{fig:waveform} for the case of a 8~mV threshold, translating into a higher jitter contribution. On the other hand, for the V$_{top}$~=~45~V, the slope of the signal is still constant in this region and this translates into a stability in the timing performance. 

In Figure~\ref{fig:Vtop45_A1_vs_A2} the time resolution measured on A1 and A2 structures, both with $V_\textrm{top}=45$~V, is compared showing a slightly better performance for the A1 version. This effect could be attributed to the different sensor layout and the increased multiplication volume in the A1. The best time resolution extracted with a fixed threshold combined with the time walk correction is $234\pm6$~ps for the A1 and $244\pm6$~ps for the A2, both achieved with a threshold of 6~mV.

The errors at 30~V in Figure~\ref{fig:timing_res} are evaluated considering the standard deviation of the time resolution of four different pixels of the same structure. Then the relative uncertainty is employed to assign an error to the 45~V results.

\newpage

\section{Investigations on gain profiles}
\label{subsec:gainprofile}

To better understand the time resolution performance of MadPix and how it was affected by the gain of the structure, we carried out further experimental and numerical analyses focusing on the gain layer.
\begin{figure}[!ht]
\centerline{\includegraphics[width=0.53\linewidth]{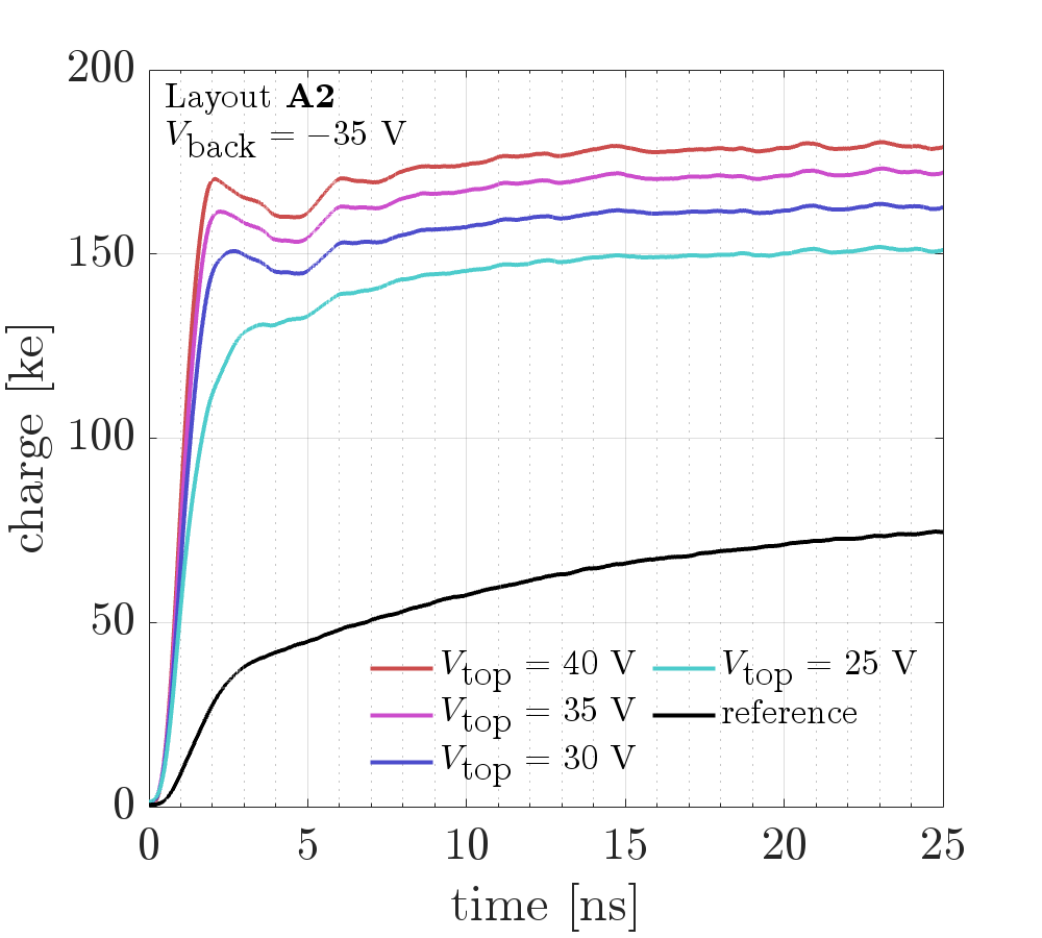}}
\caption{Measured collected charge versus time in a test-structure obtained with a 1060~nm IR pulsed laser. The curves at different top voltage refer to a device with internal gain while the black curve has been acquired with a device without gain (reference sensor).}
\label{fig:qt}
\end{figure}
In Figure~\ref{fig:qt}, several measurements of the charge collection dynamics, performed with a near-IR laser setup on test structures with and without (the reference) gain implant, are reported.
Dividing the charge obtained in devices with gain by the one acquired by the reference structure, a multiplication factor of about 2.5 was obtained.
Moreover, this plot shows that the collected charge is just slightly increasing with the top voltage, thus providing further indication of low gain.

In order to investigate the properties of the gain layer in the ARCADIA third run, we designed a numerical study based on a 2D TCAD (Technology Computer-Aided Design) framework.
Since from the capacitance variation as a function of the applied bias voltage it is generally possible to extract useful information about the multiplication implant of an LGAD diode, we proceeded to use the $C(V)$ characteristics as a benchmark tool to validate our gain calculations.
The first check that has been performed is the simulation of the lateral capacitance as a function of the top bias in a 250~$\times$~250~$\mu$m$^2$ pitch test-structure with layout A1.
Unlike in standard LGADs, we cannot infer any detail about the gain layer through a $C(V)$ obtained with a bias scan performed at the bottom.
This occurs because the \textit{p}$^+$ region on the chip backside extends throughout the whole device, and the evolution of the space charge region with bias is inadequate for probing the gain layer depletion alone.
For such reason, both measured and simulated $C(V)$ are obtained by increasing the potential between the deep-\textit{p}-well and the \textit{n}$^+$ electrode on the top.
This is why we refer to our curves as lateral $C(V)$ characteristics.
\begin{figure}[!ht]
\centering
     \begin{subfigure}[!ht]{0.49\linewidth}
         \centering
         \includegraphics[width=\textwidth]{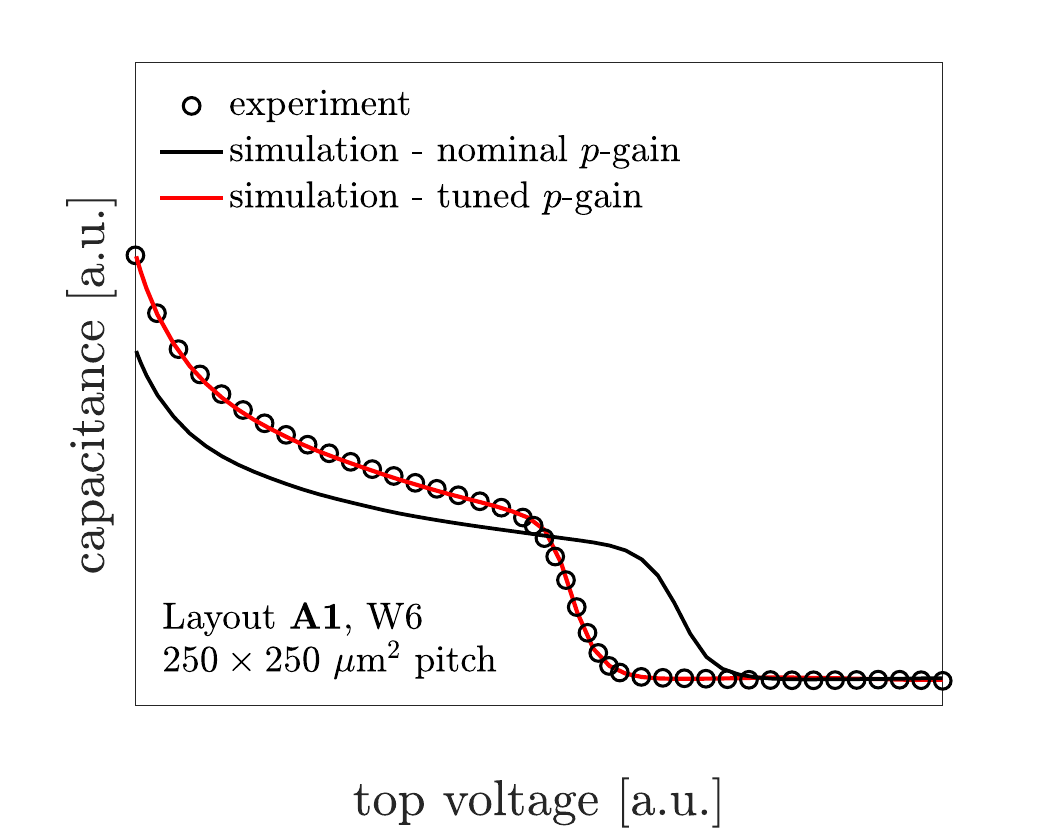}
         \caption{}
         \label{fig:CV}
     \end{subfigure}
     \begin{subfigure}[!ht]{0.49\linewidth}
         \centering
         \includegraphics[width=\textwidth]{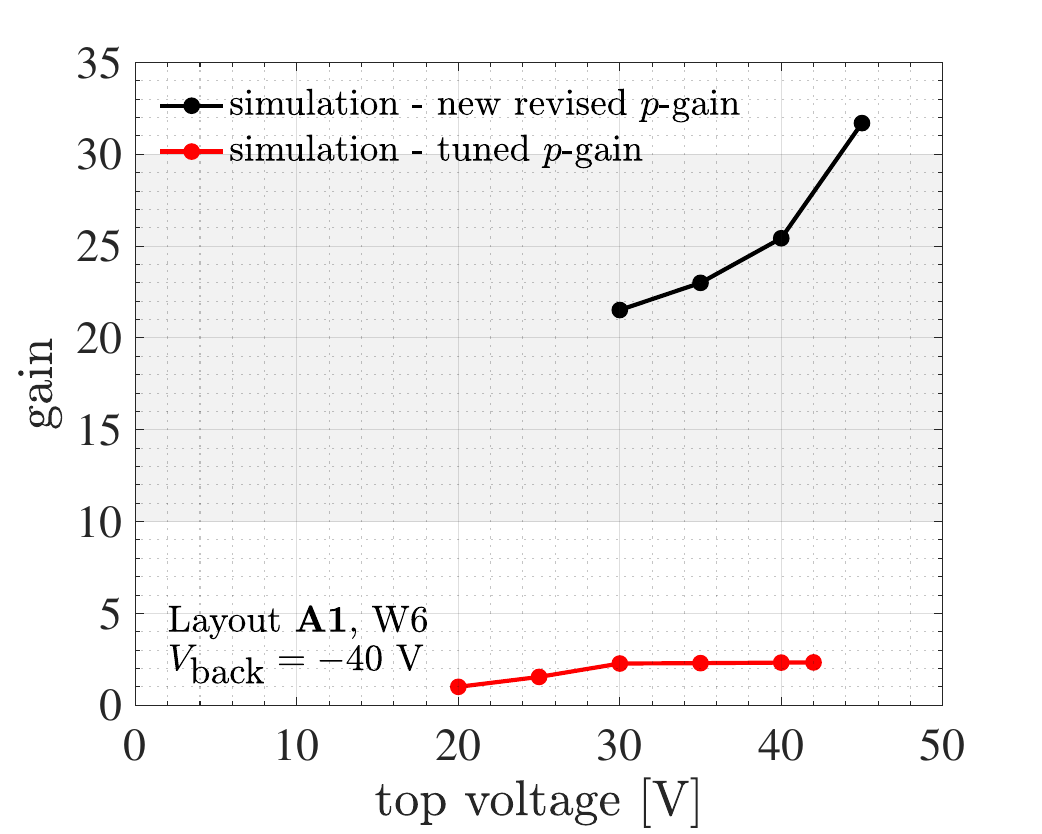}
         \caption{}
         \label{fig:GV}
     \end{subfigure}
\caption{(a) Simulated (lines) and measured (circles) lateral $C(V)$ characteristics of a 250~$\times$~250~$\mu$m$^2$ pitch test-structure with gain (units are not explicitly indicated due to non-disclosure reasons). (b) Gain curves simulated using the \textit{p}-gain implant tuned on the $C(V)$ measurements (red) and with the new profile submitted for the next short-loop run (black).}
\end{figure}
As one may see in Figure~\ref{fig:CV}, the capacitance simulated with the use of the nominal profile expected for wafer 6 (black-solid line) does not allow to reproduce the experimental $C(V)$.
In order to find a good matching between measurements and simulations, the gain implant needs to be considerably shifted towards the chip frontside (red-solid).
This fact, since the shallower the gain layer the lower the multiplication factor, could justify the gain 2.5 we measured.

To prove such hypothesis, we designed a second simulation that calculates the gain $G$ as a function of the top voltage using the tuned implantation profile as input parameter.
In the simulations, charge particles incident perpendicularly to the sensor have been considered, and the gain has been calculated as the ratio between the integrated charge with and without the multiplication model activated.
In the estimation of avalanche gain we used  van Overstraeten~-~de Man ionization coefficients~\cite{1970vanOverstraeten_SSE} and the resulting $G(V)$ curve for a device with the same layout of the previous simulations can be observed in Figure~\ref{fig:GV}.
As specified in the plot, the backside bias is set to $-$40~V, while the top voltage is sweeping in the range 20~V to 45~V.
In agreement with the results we have shown in Figure~\ref{fig:qt}, the gain simulated using the multiplication layer tuned on the measured $C(V)$ is in the range between 2 and 3, i.e. considerably lower than our expectations ($G\sim$~10-30).
This could indicate that an issue has probably occurred during fabrication.
As a matter of fact, a cross-check with the process flow revealed that the ion implantation was set at an energy 30\% lower than the agreed value.
After this check, we submitted a new short-loop run, using the same maskset of the previous production and with a revised version of the \textit{p}-gain.
This new profile should reasonably allow to obtain the initial target $G\sim$~10-30, as the simulation represented by the black curve is showing in Figure~\ref{fig:GV}.

\newpage
\section{Conclusions}
A first monolithic pixel sensor prototype with internal gain in a 110~nm CMOS platform has been produced and characterised in laboratory and in a test beam. The prototype is composed by 8 matrices with 64 pixels each. Every channel has an in pixel front-end followed by an analogue buffer outside the matrix. Two layout of the sensor terminations (A1 and A2) were implemented and tested. 
In laboratory tests the operation of each pixel was verified and the different bias points of the sensor were analysed.
With a voltage of $-30$~V on the backside, allowing the sensor to be completely depleted, and 45~V applied to the collection electrode we measured a gain of $\sim2.5$ with an IR pulsed laser.
The optical characterisation also provided an evaluation of the electronics jitter which is $\sim150$~ps for a signal amplitude of 20~mV and below 50~ps for signals above 60~mV.
The test beam campaign at CERN Proton Synchrotron with protons and pions of 10~GeV/c validated the sensor gain of $\sim2.5$ previously estimated with the IR laser and yielded the first estimate of time resolution. The A1 layout showed slightly better performance than the other one and a resolution of $234\pm6$~ps was achieved using an offline analysis with a constant 6~mV threshold and a time-walk correction based on the signal amplitude.
Furthermore, TCAD simulations on lateral capacitance of passive structures were employed to explain the measured gain around 2.5. In agreement with the foundry, a short-loop run has been scheduled and the new structures with higher gain are expected to be ready for testing in mid 2024.
Further productions will be designed with an improved front-end electronics characterised by a lower jitter to achieve the target of 20~ps total time resolution. 

\acknowledgments
This project has received funding from the European Union's Horizon 2020 Research and Innovation programme under GA no 101004761 and from the Italian National Institute for Nuclear Physics within the CSN5 call ARCADIA.
The authors gratefully acknowledge CERN and the PS staff for continuous supports to the users.

\bibliographystyle{JHEP}
\bibliography{biblio.bib}

\end{document}